\documentclass[twocolumn,eqsecnum,showpacs]{revtex4}
\usepackage[dvips]{graphicx}

\begin{document}
\title{Transition in the pattern of cracks resulting from memory effects
 in paste}

\author{Akio Nakahara}
\email{nakahara@phys.ge.cst.nihon-u.ac.jp}
\author{Yousuke Matsuo}
\affiliation{
Laboratory of Physics, College of Science and Technology \\
Nihon University, Funabashi, Chiba 274-8501, Japan
}%

\date{\today}

\begin{abstract}
\ \\
Experiments involving vibrating pastes before drying
 were performed for the purposes of controlling the crack patterns
 that appear in the drying process.
These experiments revealed a transition in the direction of lamellar cracks
 from perpendicular to parallel
 when compared with the direction of the initial external vibration
 as the solid volume fraction of the paste is decreased.
This result suggests a transition in the "memory" in paste,
 which is visually represented as morphological changes in the crack pattern.
Given that the memory in paste represents the flow pattern
 induced by the initial external vibration,
 it should then be possible to control and design various crack patterns,
 such as cellular, lamellar, radial, ring, spiral, and others.
\end{abstract}

\maketitle


It is very important to find methods to control crack patterns
 in various fields of science and technology.
So far, few successful examples of such control have been reported,
 one such success being the case of quasi-static fractures
 in thin glass plate
 in which control of the morphology of crack patterns
 was proven to be possible
 by changing the cooling rate
 during quenching~\cite{Yuse93, Marder94, Hayakawa94, Sasa94}.
Recently,
 we have developed a method for controlling the morphology of
 the crack patterns that appear
 during the drying of pastes~\cite{Nakahara05a, Nakahara05b, Nakahara06}.
A paste is a concentrated colloidal suspension with plasticity.
When a mechanical force is applied to a paste before drying,
 the paste "remembers" the force
 in the form of microscopic structural changes within the suspension.
This "memory" is represented visually by the crack pattern
 that later appears in the drying process.
In this Rapid Communication,
 anisotropic colloidal particles were used to produce
 a transition in the crack patterns of pastes,
 which suggests the existence of various memory structures in the suspension.

A paste, as defined in this study,
 consists of a powder mixed with water.
This mixture is poured into a container
 and keep in an air-conditioned room
 at a fixed temperature of 25${\rm {}^o C}$ and a humidity of 30\%.
As the mixture dries, the mixture shrinks due to the evaporation of the water.
When
 adhesion between the mixture and the container is strong,
 the mixture sticks to the bottom of the container,
 prohibiting contraction of the mixture in the horizontal direction
 unless the mixture splits into smaller pieces through the formation of cracks.
The formation of cracks continues
 until the sizes of the fragments are approximately the same as the thickness
 of the mixture~\cite{Groisman94, Allain95, Komatsu97, Kitsune99}.
Usually, crack patterns form isotropic cellular structures,
 in that
 the sizes of these fragments are almost the same,
 the shape of each fragment is isotropic,
 and they are distributed randomly in space.

Recent work has indicated that
 the morphology of crack patterns can be controlled
 by applying an external horizontal vibration to a paste
 before it has dried~\cite{Nakahara05a, Nakahara05b, Nakahara06, Ooshida05, Otsuki05}.
In these experiments,
 we used calcium carbonate particles ${\rm CaCO_3}$ as the powder to make pastes.
Applying an initial vibration to the paste
 with a strength just above that of the yield stress of the paste
 led to the formation of lamellar cracks during the drying process.
The orientation of these lamellar cracks was
 perpendicular to the direction of the initial vibration.

In this Rapid Communication,
 the powder was changed from calcium carbonate
 to magnesium carbonate hydroxide (Kanto Chemical, Tokyo, Japan),
 and the same experiments were performed as those described above.
The density of magnesium carbonate hydroxide is ${\rm 2.0 g/cm^3}$.
Microscopic observations
 via a scanning electron microscope (Hitachi, Tokyo, Japan) indicated
 that the shapes of dry magnesium carbonate hydroxide particles are disk-like,
 with a diameter of ${\rm 1.5 \mu m}$ and a thickness of ${\rm 0.2 \mu m}$.
In these experiments,
 square containers were used with the length of each side being 200 mm.
The mass of the powder in the mixture was fixed at 100 g in each container,
 such that the final thickness of the mixtures with different solid volume fractions
 was equal when they dried,
 thus equalizing the characteristic sizes of the final crack patterns.
The pastes were vibrated horizontally for 60 s
 immediately after pouring the paste into the container.
The formation of cracks was then observed as the pastes dried.

\begin{figure}
\includegraphics*[width=8.6cm]{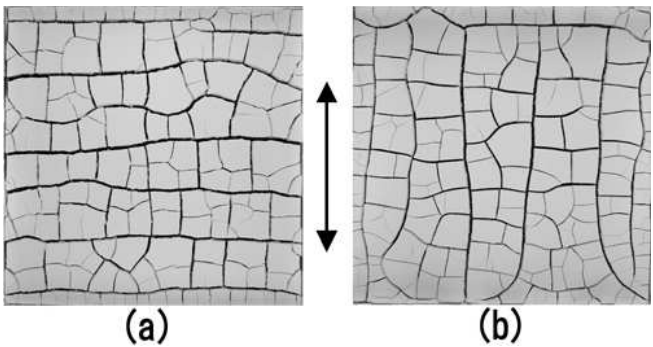}
\caption{Transition in the direction of lamellar crack patterns.
The arrow between (a) and (b) indicates the direction of the initial vibration,
 where the amplitude $r$ and the frequency $f$ of the vibration are 15 mm and 2 Hz,
 respectively, i.e., the strength $4 \pi^2 r f^2$ is 2.4${\rm m/s^2}$.
The lengths of the sides of both square containers
 are 200 mm.
 (a) The solid volume fraction $\rho$ is 12.5\%.
The direction of the lamellar cracks is perpendicular
 to the direction of the initial vibration.
 (b) The solid volume fraction $\rho$ is 6.7\%.
The direction of the lamellar cracks is parallel
 to the direction of the initial vibration.
}
\label{eps1}
\end{figure}

\begin{figure}
\includegraphics*[width=8cm]{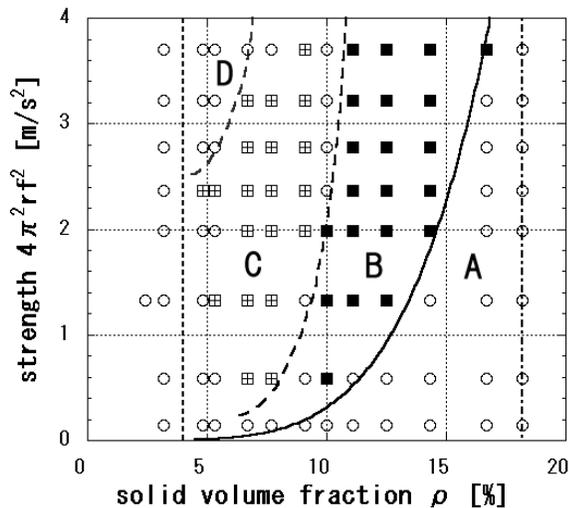}
\caption{Morphological phase diagram of the crack patterns
 that appear in the drying process of pastes
 made of magnesium carbonate hydroxide,
 shown as a function of the solid volume fraction $\rho$
 and the strength $4 \pi^2 r f^2$ of the initial vibration.
Open circles denote isotropic cellular crack patterns,
 solid squares denote lamellar crack patterns,
 the direction of which is perpendicular
 to the direction of the initial vibration,
 and open squares with a plus inside denote lamellar crack patterns,
 the direction of which is parallel
 to the direction of the initial vibration.
}
\label{eps2}
\end{figure}

Figure~\ref{eps1} indicates that
 a decrease in the solid volume fraction $\rho$ of the paste
 results in a transition in the direction of the lamellar crack patterns
 from perpendicular (Fig.1(a)) to parallel (Fig.1(b))
 when compared with the direction of the initial vibration.
Systematic experiments were performed
 to obtain a morphological phase diagram of the crack patterns,
 shown in Fig.~\ref{eps2},
 as a function of both the solid volume fraction $\rho$
 and the strength $4 \pi^2 r f^2$ of the initial vibration.
We also performed rheological measurements of the yield stress of the pastes
 via a Dynamic Stress Rheometer (Rheometrics, Piscataway, NJ).
When the value of $\rho$ is smaller than the Liquid-Limit $\rho=4.0\%$,
 which is denoted by the vertical dotted line,
 only isotropic cellular crack patterns appear.
When the value of $\rho$ is larger than the Plastic-Limit $\rho=18.2\%$,
 which is denoted by the vertical dashed-and-dotted line,
 a morphological phase diagram is not obtained,
 as mixing powder with water homogeneously is not possible
 due to the lack of a sufficient amount of water.
The region between these two vertical lines is divided
 by the solid and the dashed curves into four regions, A, B, C, and D.
The solid curve corresponds to the yield stress line,
 on which the strength of the initial vibration is
 equal to the strength of the yield stress of the paste.
Isotropic cellular crack patterns appear in regions A and D.
In region A, the value of the strength $4 \pi^2 r f^2$
 of the initial vibration is smaller than that of the yield stress,
 while in region D, turbulent flow induced by the initial vibration
 destroys the memory.
In regions B and C,
 lamellar crack patterns are obtained
 that indicate a memory of the initial vibration,
 but the directions of the lamellar cracks are
 perpendicular in region B and parallel in region C
 when compared with the direction of the initial vibration.

We previously reported that
 in the drying process of a paste made of calcium carbonate,
 lamellar cracks emerged when the paste was vibrated
 at the strength just above the yield stress,
 and the directions of the lamellar cracks were all perpendicular to the direction
 of the initial external vibration~\cite{Nakahara05a, Nakahara05b, Nakahara06}.
This corresponds to the formation of lamellar cracks in region B in Fig.~\ref{eps2}.
It is postulated that,
 when the densely packed colloidal particles are vibrated,
 longitudinal density fluctuations emerge
 inside the suspension due to the inelastic collisions between particles.
This is similar to the formation of density waves or jammed structures
 in dense granular flows~\cite{Nakahara97}.
The polydispersity of the particles also helps
 the shear-induced segregation caused by horizontal oscillations
 to form a lamellar pattern perpendicular
 to the direction of the oscillation~\cite{Ciamarra05}.
Due to its plasticity,
 the network of colloidal particles maintains the microscopic anisotropic structure
 induced by the initial external vibration
 even after the vibration is removed~\cite{Vanel99}.

\begin{figure}
\includegraphics*[width=7cm]{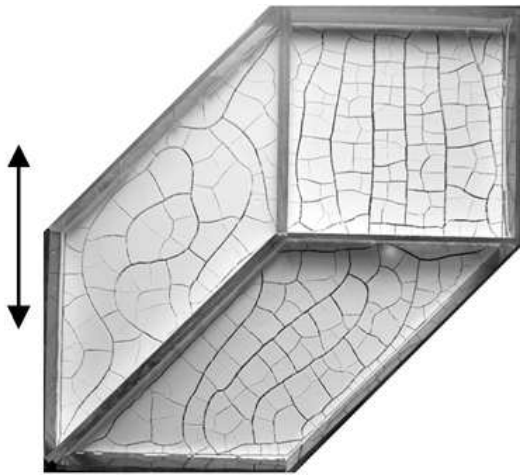}
\caption{Memory of flow patterns in region C.
Here,
 the solid volume fraction
 $\rho$ is 7.7\%.
The length of each side of the square container is 200 mm
 and the lengths of the neighboring sides
 of the two parallelogrammatic containers are
 200 mm and ${\rm 200 \sqrt{2}}$ mm.
The arrow indicates the direction of the initial external vibration,
 where the amplitude $r$ and the frequency $f$
 of the vibration are 15 mm and 1 Hz, respectively,
 i.e., the strength $4 \pi^2 r f^2$ is 0.6${\rm m/s^2}$.
The flow patterns caused by the initial external vibration
 are memorized and are exhibited by the crack patterns.
}
\label{eps3}
\end{figure}

As for the lamellar cracks in region C in Fig.~\ref{eps2}, however,
 it is believed that there is a relation
 between the direction of the lamellar cracks
 and the direction of the flow induced by the initial external vibration.
To test this theory,
 experiments were performed in different container shapes.
Figure~\ref{eps3} shows that,
 in parallelogrammatic containers, the direction of lamellar cracks is
 parallel to the direction of the flow along the oblique direction
 and not to the direction of the external vibration.
That is, the memory in region C represents the flow pattern.

\begin{figure}
\includegraphics*[width=6.5cm]{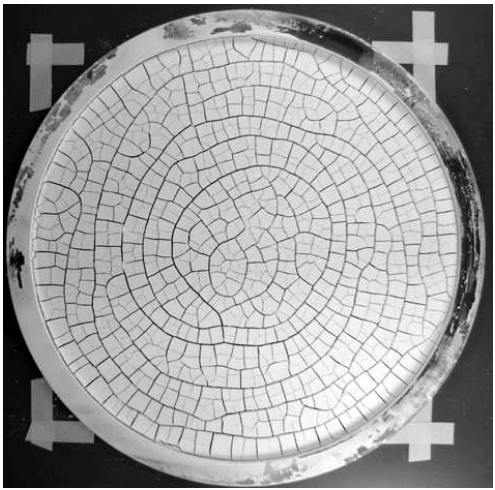}
\caption{A ring pattern.
Here,
 the solid volume fraction
 $\rho$ is 7.7\%.
The diameter of the circular container is 500 mm.
Before drying, we put a cover on the paste,
 rotated the cover counterclockwise at a period of 1 s for 10 s,
 and then removed the cover for drying.
It can be seen that no surface wave or vibration is necessary
 for the formation of a flow pattern.
}
\label{eps4}
\end{figure}

Figure~\ref{eps4} shows the results of experiments designed
 to check the effects of surface waves induced by the initial vibration.
Here, the solid volume fraction $\rho$ is 7.7\%,
 which corresponds to the paste in region C.
We used a circular container with a diameter of 500 mm,
 and the mass of the powder in the mixture was 491 g
 in order to equal the final thickness of the mixtures
 in Figs.~\ref{eps1}-\ref{eps3}.
Before drying,
 the paste was covered to suppress the formation of surface waves.
The cover was then rotated counterclockwise, and then removed for drying.
Figure~\ref{eps4} shows that even with no surface waves
 a ring pattern forms,
 which is the flow pattern produced by the rotation of the cover.
It also shows that
 vibration is not necessary for the formation of
 a ring pattern.
Note that vibration is necessary
 for the formation of lamellar cracks in region B.
The transition of the direction of the lamellar cracks from region B to region C
 is a visual representation of the transition of the memory in the paste
 from "memory of vibration" to "memory of flow".

\begin{figure}
\includegraphics*[width=5cm]{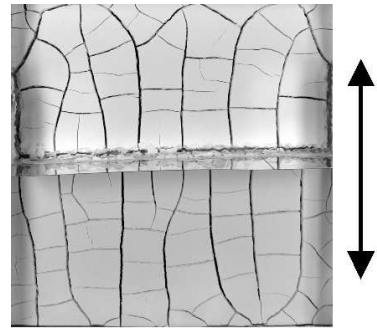}
\caption{Experimental result
 to check the position of the lamellar cracks in region C.
Here,
 the solid volume fraction
 $\rho$ is 6.7\%,
 and the lengths of the sides of the square container are 200 mm.
The arrow indicates the direction of the initial vibration,
 where the amplitude $r$ and the frequency $f$ of the vibration are 15 mm and 1.5 Hz,
 respectively,  i.e., the strength $4 \pi^2 r f^2$ is 1.3${\rm m/s^2}$.
After vibrating the paste, we cut the paste into halves and let them dry.
The positions of the lamellar cracks
 are discontinuous at the boundary between the halves.
}
\label{eps5}
\end{figure}

Given the above results,
 one question that can be asked is,
 why does the paste made of magnesium carbonate hydroxide
 retain the flow pattern,
 while that of calcium carbonate does not?
To answer this,
 we must consider how the particles interact
 with each other in each of the suspension.
In both pastes,
 the colloidal particles interact via van der Waals attractive forces.
Additionally, ${\rm CaCO_3}$ particles also interact with each other
 through Coulombic repulsive forces~\cite{Nakahara06}.
Given these Coulombic repulsive forces,
 only at high solid volume fractions
 at which particles are densely packed
 do ${\rm CaCO_3}$ particles attract each other to form a network structure.
A mixture of powder and water with this high of a solid fraction
 has plasticity with a nonzero yield stress.
In comparison,
 the colloidal suspension of magnesium carbonate hydroxide
 can form a network structure
 even at low solid volume fractions,
 because the particles interact via attractive forces only.
The shapes of the particles also play a role.
It is easy to pack ${\rm CaCO_3}$ particles densely
 because the shape of the particles is isotropic,
 while the disk-like particles of magnesium carbonate hydroxide stick together easily
 and form a structure similar to a card house with many voids in the interior,
 resulting in a low solid volume fraction.
This rationale is supported by the experimental results in that
 the Liquid-Limit and the Plastic-Limit
 of the colloidal suspensions of calcium carbonate
 were as high as 25.0\% and 54.0\%, respectively,
 while those of magnesium carbonate hydroxide were as low as 4.0\% and 18.2\%,
 respectively.
Since a colloidal suspension of magnesium carbonate hydroxide has plasticity
 even at low solid volume fractions,
 the flow that emerges at low solid volume fractions can be retained.
It is believed that the flow induced by the initial external vibration
 elongates a dilute network structure to produce a lamellar microstructure,
 the direction of which is parallel to the direction of the flow,
 with the normals of the disk-like particles
 perpendicular to the direction of the flow~\cite{Dzubiella02, Vermant05}.
In fact,
 when we prepared two pastes with the same solid volume fraction
 and vibrated only one of them before drying,
 the lamellar cracks that appeared in the vibrated paste
 always emerged earlier than the isotropic cellular cracks
 in the paste that was not vibrated.
This result indicates that
 the network of colloidal particles is weakened
 by the formation of an anisotropic microstructure.

It is believed that the order parameter
 that quantifies the transition between phases B and C
 is anisotropy in the microstructure,
 which is visualized by the macroscopic crack pattern.
Only one shake is enough to create anisotropy in a paste,
but to obtain a beautiful lamellar crack pattern,
 a paste should be vibrated more than 10 times.
Note that near the transition line between phases B and C,
 a combined lamellar and cellular pattern is observed.
Since the level of anisotropy is low near the transition line,
 dynamic instability results in unstable straight crack propagation,
 which leads to a cellular pattern in crack formation~\cite{Nakahara06}.

\begin{figure}
\includegraphics*[width=6.5cm]{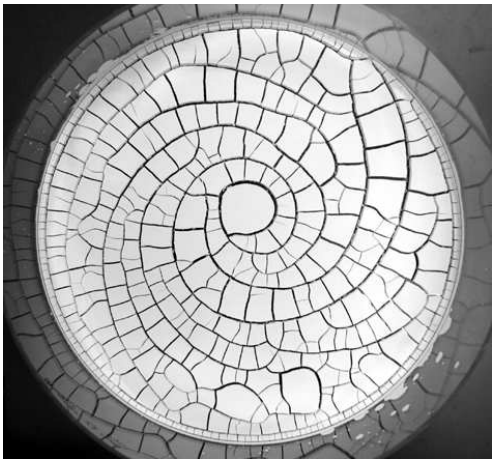}
\caption{A spiral crack pattern.
Here,
 the solid volume fraction
 $\rho$ is 7.7\%.
The diameter of the circular container is 500 mm.
Before drying,
 the container was translated along a circuit
 with a radius of 15 mm clockwise at a period of 1 s for 60 s,
 so that the container pushes the paste
 inward and clockwise simultaneously to produce a spiral flow.
Note that the container was not rotated
 around its center.
}
\label{eps6}
\end{figure}

The positions of the lamellar cracks in region C were checked
 to explore the possibility
 that the lamellar cracks were induced by macroscopic hydrodynamic instability.
If the positions of the lamellar cracks were determined by the initial vibration,
 macroscopic hydrodynamic instability,
 such as convection or surface waves,
 should play a role in memory formation in pastes.
Any effects from surface waves can be eliminated
 based on the results in Fig.~\ref{eps4}.
Figure~\ref{eps5} shows that
 even if the directions of the lamellar cracks are determined
 by the direction of the initial vibration,
 the positions of the lamellar cracks are random.
These experimental results support our interpretation
 that the memory in pastes is retained by a microscopic anisotropic structure
 of an orientational order,
 the scale of which is extremely small
 compared with the characteristic size of the macroscopic crack patterns.

To summarize,
 we performed drying experiments and revealed
 a transition in the memory in pastes
 that is visually represented as morphological changes in the crack pattern.
The ability to mechanically imprint a flow pattern into a paste
 by changing the shape of container and/or the mechanical force
 allows for design of various crack patterns,
 including cellular, lamellar, radial~\cite{Nakahara05a, Nakahara05b, Nakahara06},
 ring, and spiral (shown in Fig.~\ref{eps6}) patterns.

We thank
 H. Uematsu, M. Otsuki, S. Sasa, T. S. Komatsu, T. Ooshida, M. K\"ulzer,
 and Y. Nakahara for their valuable discussions.
We also thank
 M. Sugimoto, T. Taniguchi, and K. Koyama
 for their support with the rheological measurements
 at Venture Business Laboratory of Yamagata University,
 and Y. Aoyagi, A. Taguchi, K. Nakagawa, and A. Itoh
 for their support with the microscopic observations
 via a Scanning Electron Microscope
 at the Advanced Materials Science Center of Nihon University.

\end{document}